\documentclass[preprint,showpacs,preprintnumbers,amsmath,amssymb]{revtex4}

\usepackage{graphicx}
\usepackage{dcolumn}
\usepackage{bm}

\begin{document}


\title{Scalar-vector Lagrangian without nonlinear self-interactions 
of bosonic fields in the relativistic mean-field theory}

\author{M.M. Sharma}
\affiliation{Physics Department, Kuwait University, Kuwait 13060}

\date{\today}

\begin{abstract}

A new Lagrangian model without nonlinear scalar self-interactions in the 
relativistic mean-field (RMF) theory is proposed. Introducing terms for 
scalar-vector interactions (SVI), we have developed a RMF Lagrangian 
model for finite nuclei and nuclear matter. It is shown that by inclusion 
of SVI in the basic RMF Lagrangian, the nonlinear $\sigma^3$ and
$\sigma^4$ terms can be dispensed with. The SVI Lagrangian thus obtained 
provides a good description of ground-state properties of nuclei 
along the stability line as well as far away from it. This Lagrangian model
is also able to describe experimental data on the breathing-mode 
giant monopole resonance energies well.

\end{abstract}




\pacs {21.30.Fe, 21.10.Dr, 21.60.-n, 24.10.Cn, 24.10.Jv, 21.65.+f}



\maketitle

The relativistic mean-field (RMF) theory \cite{SW.86,Rein.89,Ser.92} 
of nuclear interaction based upon exchange of mesons between nucleons
has established itself as a successful approach to describing properties 
of nuclei along the stability line as well as far away from 
it \cite{GRT.90,SNR.93,Lala.97}. The Dirac-Lorentz structure of 
nucleons provides a built-in spin-orbit interaction with its advantage 
over the non-relativistic Skyrme theory in describing properties 
such as anomalous isotope shifts in Pb nuclei \cite{SLR.93}, which depend
upon shell structural effects. An isospin dependence of the spin-orbit 
interaction or rather a lack of it is responsible for the anomalous 
behaviour of the isotope shifts \cite{SLK.95}. The idea of pseudospin 
symmetry in nuclei has been attributed to the relativistic (Dirac) 
nature of nucleons \cite{ginocchio.97}. 

With the advent of the Walecka model \cite{SW.86} for nuclei and nuclear
matter, it was realized that the linear RMF Lagrangian with its 
large value of nuclear incompressibility was unable to describe properties
of finite nuclei properly. A lack of proper ingredients for a suitable 
description of surface properties was cited as a main reason for this drawback.
In order to remedy this problem, Boguta and Bodmer \cite{BB.77} introduced 
nonlinear scalar self-coupling terms of the form $\sigma^3$ + $\sigma^4$. 
Consequently, the RMF Lagrangian has proved to be successful and the nonlinear 
scalar terms have thus become indispensable for an adequate treatment of
finite nuclei and nuclear matter saturation. The nonlinear $\sigma$-field 
seems to provide an essential density dependence of nuclear force in a 
finite nucleus. With the inclusion of the nonlinear $\sigma$-field, the theory
also becomes renormalizable. 

One of the first successful nuclear forces within this model is 
NL-SH \cite{SNR.93}. Within this model an improved
set NL3 has been obtained for finite nuclei \cite{Lala.97}. 
However, as is well known, the model with $\sigma^3$ + $\sigma^4$ 
gives an equation of state (EOS) of nuclear matter 
that is very stiff and is consequently untenable for the 
spectrum of observed neutron star masses. 

The quartic vector coupling of the form $\omega^4$  was introduced 
\cite{Bodmer.89} in the RMF Lagrangian. An appropriate
description of finite nuclei was obtained with the vector
self-coupling of $\omega$ meson \cite{Toki.98,Sharma.00}. 
The inclusion of the $\omega^4$ coupling has also helped to 
improve the shell effects along the stability line \cite{Sharma.00}. 
A desired softening of the EOS of nuclear matter due to vector 
self-coupling of $\omega$ meson was shown in ref. \cite{Mueller.96}. 

The density dependence of the nuclear interaction in the RMF theory
remains an open problem.  The Walecka model offers an appropriate
avenue to construct a theory that should be suitable for describing
various aspects of finite nuclei all along the periodic table
(an ambitious goal) as well as properties of nuclear matter 
concomitantly. Point-coupling models have been introduced 
\cite{Nikolaus.92,Buerv.02} to describe finite nuclei.
Attempts have been made to broaden the basis of the RMF Lagrangian 
by including terms of higher orders in the scalar and vector fields 
with inclusion of interaction terms amongst various mesonic 
degrees of freedom \cite{Savush.97,Furn.97,Todd.05}. Density-dependent 
meson couplings \cite{Brock.92,Fuchs.95,Typel.05,Nisk.02,Long.04} have been 
introduced with a view to modify density dependence of the nuclear
interaction in an explicit form with a good degree of success. This requires
inclusion of additional parameters to model density-dependence of meson
couplings. Notwithstanding the above, the RMF theory serves as an 
ideal platform for an effective field theoretical approach for 
many-body problems of nuclei with sufficient space for innovation. 

An upsurge in experimental data especially in the domain of extreme
regions of the periodic table provides an incentive to devise new and 
improved approaches and models to be able to describe the same. 
Savushkin et al. \cite{Savush.97} have incorporated various
meson-meson interactions in their approach especially those between
$\sigma$ and $\omega$ meson in addition to nonlinear couplings of both
these mesonic fields. This problem has been approached \cite{Furn.97}
from a more general point of view by taking expansion 
in and interactions amongst various mesonic fields. This approach
has led to an improvement for finite nuclei and nuclear matter
with a larger number of parameters required.

In this work, we have sought to explore the possibility of dispensing
with the nonlinear scalar self-couplings which have so far remained 
essential for finite nuclei.  We ask ourselves: whether it is possible 
to mock the scalar self-couplings and their inherent density dependence 
in nuclei by employing meson-meson interactions especially between 
$\sigma$ and $\omega$ mesons instead? Keeping the issue of renormalizability 
in abeyance, we have added couplings between $\sigma$ and 
$\omega$ mesons of the form  $\sigma \omega^2$ + $\sigma^2 \omega^2$ to the 
basic (linear) RMF Lagrangian based upon exchange of $\sigma$, $\omega$ and
$\rho$ mesons. Properties of nuclear matter for the scalar-vector interaction 
(SVI) of this form were explored in ref. \cite{Moncada.94}. 
Recently, the Lagrangian model SIG-OM with the 
inclusion of the coupling of the form $\sigma^2 \omega^2$ 
whilst retaining the scalar self-couplings $\sigma^3$ + $\sigma^4$ has 
been developed \cite{Haidari.07}. In the present work, we have narrowed 
down the space by excluding the self-couplings at the expense of 
the scalar-vector meson-meson couplings.

The basic RMF Lagrangian density that describes nucleons as Dirac 
spinors interacting with the meson fields is given by \cite{SW.86}
\begin{eqnarray}
{\cal L}_0&=& \bar\psi \left( \rlap{/}p - g_\omega\rlap{/}\omega -
g_\rho\rlap{/}\vec\rho\vec\tau - \frac{1}{2}e(1 - \tau_3)\rlap{\,/}A -
g_\sigma\sigma - M_N\right)\psi\nonumber\\
&&+\frac{1}{2}\partial_\mu\sigma\partial^\mu\sigma- \frac{1}{2} m^2_\sigma \sigma^2_{}
-\frac{1}{4}\Omega_{\mu\nu}\Omega^{\mu\nu}+ \frac{1}{2}\nonumber
m^2_\omega\omega_\mu\omega^\mu\\ &&
-\frac{1}{4}\vec R_{\mu\nu}\vec R^{\mu\nu}+
\frac{1}{2} m^2_\rho\vec\rho_\mu\vec\rho^\mu -\frac{1}{4}F_{\mu\nu}F^{\mu\nu},
\label{eq:one}
\end{eqnarray}
where $M_N$ is the bare nucleon mass and $\psi$ is its Dirac spinor. 
Nucleons interact with $\sigma$, $\omega$, and $\rho$ mesons, 
with coupling constants being $g_\sigma$, $g_\omega$ and $g_\rho$, 
respectively. The photonic field is represented by the electromagnetic 
vector $A^\mu$. The effective Lagrangian for finite nuclei that is used
commonly is given by
\begin{eqnarray}
{\cal L}_{\rm eff} = {\cal L}_0 - U_{\rm NL}. 
\end{eqnarray}
The nonlinear $\sigma$-meson self-couplings which have so far been 
an integral part of the RMF Lagrangian are given by
\begin{equation}
U_{\rm NL} = \frac{1}{3}g_2\sigma^3_{} + \frac{1}{4}g_3\sigma^4. 
\end{equation}
The parameters $g_2$ and $g_3$ are the nonlinear couplings of $\sigma$-meson
in the conventional $\sigma^3$ + $\sigma^4$ model \cite{BB.77}. Here, we put 
$g_2 = g_3 = 0$, thus eliminating the self-couplings  $U_{\rm NL}$ of $\sigma$
meson. Instead, we introduce the meson-meson interaction terms of the form
\begin{eqnarray}
U_{\rm mm} = \frac{1}{2}g_4 \sigma \omega_{\mu}\omega^{\mu}  + 
 \frac{1}{2}g_5 \sigma^2 \omega_{\mu}\omega^{\mu}.
\end{eqnarray}
where $g_4$ and $g_5$ represent the respective coupling constants for
meson-meson interactions between $\sigma$ and $\omega$ mesons. The effective
Lagrangian in our case is then
\begin{eqnarray}
{\cal L}_{\rm eff} = {\cal L}_0 + U_{\rm mm}. 
\end{eqnarray}
The corresponding Klein-Gordon equations can be written as
\begin{eqnarray}
( -\Delta + m_{\sigma}^{*2} ) \sigma & = & -g_\sigma\bar\psi\psi \nonumber\\
( -\Delta + m_{\omega}^{*2} ) \omega_\nu & = 
          & g_\omega\bar\psi\gamma_\nu\psi \nonumber \\
( -\Delta + m_{\rho}^{2} ) \vec\rho_\nu & = & g_{\rho}
\bar\psi\gamma_\nu\vec\tau\psi \nonumber \\
 -\Delta A_\nu & = & \frac{1}{2}e\bar\psi(1 + \tau_3)\gamma_\nu\psi,
\end{eqnarray}
where the effective meson masses $m_{\sigma}^*$ and $m_{\omega}^*$ can be
obtained as
\begin{eqnarray}
m_{\sigma}^{*2} & = & m_{\sigma}^2  - g_4{\omega_0}^2/(2\sigma) 
               -  g_5{\omega_0}^2 \nonumber\\
m_{\omega}^{*2} & = & m_{\omega}^2 + g_4 \sigma  + g_5 \sigma^2.
\end{eqnarray}
These equations represent an implicit density dependence of $\sigma$ and $\omega$
meson masses and effectively that of the nuclear interaction therein.

The parameters of the new Lagrangian model SVI are obtained by a multi-dimensional
search in the parameter space by fitting experimental binding energies and charge 
radii of a set of a few nuclei (cf. \cite{SNR.93} for a detailed
procedure). The nuclei included are $^{16}$O, $^{40}$Ca, $^{90}$Zr,
$^{116}$Sn, $^{124}$Sn and $^{208}$Pb. The isotopes $^{116}$Sn and $^{124}$Sn 
are included in order to span the broad range of isospin. No conditions have 
been put on nuclear matter properties and thus the parameters
are allowed to vary freely without any bias to the nuclear matter properties.
The $\omega$ and $\rho$ meson masses have been fixed at their empirical values. 

\begin{table}
\caption{\label{tab:table1}The parameters and nuclear matter (NM) properties of 
the scalar-vector Lagrangians SVI-1 and SVI-2 without nonlinear scalar 
self-couplings. The sets NL-SH and NL3 with the scalar self-couplings 
are also shown for comparison.}
\begin{ruledtabular}
\begin{tabular}{l l l l l l}
& Parameters        & SVI-1   & SVI-2   & NL-SH     &  NL3 \\    
\hline 
&M (MeV)            & 939.0   & 9.39.0  & 939.0     & 939.0    \\
&$m_{\sigma}$ (MeV) & 524.527  & 524.024  & 526.0592  & 508.194 \\
&$m_{\omega}$ (MeV) & 783.0   & 783.0    & 783.0     & 782.501  \\
&$m_{\rho}$ (MeV)   & 763.0   & 763.0    & 763.0     & 763.0    \\
&$g_{\sigma}$       & 9.6762  & 9.641    & 10.4436   &  10.217 \\
&$g_{\omega}$       & 11.6028 & 11.565   & 12.9451   &  12.8675 \\
&$g_{\rho}$         & 4.464  & 4.492    & 4.3828    &   4.4744 \\
&$g_2$ (fm$^{-1}$)  &  0      & 0        & $-$6.9099  & $-$10.432 \\
&$g_3$              &  0      & 0        & $-$15.8337 & $-$28.885 \\
&$g_4$ (fm$^{-1}$)  & 17.1537 & 16.962   & ~~0.0      &  ~~0.0   \\ 
&$g_5$              & 33.8565 & 32.819   & ~~0.0      &  ~~0.0   \\ \\
& NM properties & & & &\\
& $\rho_0$ (fm$^{-3}$) & 0.149    & 0.149  & 0.146      & 0.148    \\
& $a_v$ (MeV)          & $-$16.30 & $-$16.31 & $-$16.33   & $-$16.24 \\
 & $K$ (MeV)           & 263.9    & 271.5  & 354.9      & 271.6    \\
& $m^*$                & 0.616    & 0.621  & 0.597      & 0.595    \\
& $J$ (MeV)          & 37.6     & 37.0   & 37.0       & 37.4     \\
\end{tabular}
\end{ruledtabular}
\vspace{-0.3cm}
\end{table}

The parameters of the Lagrangian obtained as a result of a free variation 
in the multi-dimensional space are shown in Table I. We have obtained 
two parameter sets SVI-1 and SVI-2 which are deemed as appropriate 
for ground-state binding energies and charge radii of nuclei.
The parameter of the forces NL-SH and NL3 with the nonlinear scalar
couplings are also shown for comparison. 

The nuclear matter properties of SVI-1 and SVI-2 are shown in the lower section 
of Table I. The sets SVI-1 and SVI-2 are close to each other in the
nuclear matter properties with a slight difference in the incompressibility 
with $K = 264$ MeV for SVI-1 and $K = 272 $ MeV for SVI-2. There is only a 
minor difference in the effective mass $m^*$. 
The $m^*$ values for SVI interactions are clearly higher than those of 
the Lagrangian sets NL-SH and NL3 with the scalar self-interactions.

The saturation density for both SVI-1 and SVI-2 is slightly higher than
that of NL-SH and NL3. One notable difference between the two SVI sets is
the difference in the asymmetry energy $J$ (or $a_4$). One can note that
even in an interaction that is different from the scalar self-interactions,
it is not possible to bring down the asymmetry energy in the acceptable 
range of 30-33 MeV.

The binding energy of key spherical nuclei along with a few representative ones
as obtained with SVI-1 and SVI-2 is shown in Table II. For a comparison,
we also show the results due to NL-SH and NL3. With the exception of the very 
light nucleus of $^{16}$O, both SVI-1 and SVI-2 show an excellent 
agreement with the experimental binding energies over a large range of mass.
This is reflected by the smaller value of the $rms$ deviation $\delta$ of
SVI-1 and SVI-2 values from the experimental data vis-a-vis NL3 and NL-SH
shown at the bottom of Table II. A marked improvement with SVI interactions 
is in the binding energy of doubly magic nuclei $^{100}$Sn, $^{132}$Sn and 
$^{208}$Pb over those of NL-SH and NL3. This may have consequences on the 
shell effects in nuclei especially in the vicinity of the r-process path 
and drip lines.

\begin{table}
\caption{\label{tab:table2}The binding energy (in MeV) of nuclei calculated 
with SVI-1 and SVI-2 and compared with NL3 and NL-SH. The empirical values (exp.) 
are shown in the last column. The $rms$ deviation ($\delta$) of theoretical 
values from the experimental data is shown at the bottom of the table.}
\begin{ruledtabular}
\begin{tabular}{l l l l l l l}& Nucleus        & ~SVI-1   & ~SVI-2  & ~NL-SH     &   ~~NL3     &  ~~exp.~~   \\    
\hline 
& $^{16}$O       & $-$129.7   &$-$129.7    & $-$128.4  & $-$128.8  & $-$127.6 \\
& $^{40}$Ca      & $-$343.2   &$-$343.2    & $-$340.1  & $-$342.0  & $-$342.1 \\
& $^{48}$Ca      & $-$415.3   &$-$415.0    & $-$415.1  & $-$415.2  & $-$416.0 \\
& $^{90}$Zr      & $-$783.0   &$-$783.0    & $-$782.9  & $-$782.6  & $-$783.9 \\
& $^{100}$Sn     & $-$827.3   &$-$827.1   & $-$830.6  & $-$829.2  & $-$824.5 \\  
& $^{116}$Sn     & $-$988.3   &$-$988.4   & $-$987.9  & $-$987.7  & $-$988.7 \\ 
& $^{124}$Sn     & $-$1049.7  &$-$1049.6  & $-$1050.1 & $-$1050.2 & $-$1050.0 \\  
& $^{132}$Sn     & $-$1103.8  &$-$1103.3  & $-$1105.9 & $-$1105.4 & $-$1102.9 \\  
& $^{202}$Pb     & $-$1591.2  &$-$1591.8   & $-$1595.8 & $-$1592.6 & $-$1592.2 \\
& $^{208}$Pb     & $-$1637.0  &$-$1637.3   & $-$1640.4 & $-$1639.5 & $-$1636.7 \\
& $^{214}$Pb     & $-$1661.7  &$-$1662.3   & $-$1664.3 & $-$1661.6 & $-$1663.3 \\
\hline
& $\delta$       & ~~~~1.33   &  ~~~~1.20  &  ~~~~2.70 &  ~~~~2.00   & \\
\end{tabular}
\end{ruledtabular}
\end{table}

The charge radii of nuclei obtained with SVI-1 and SVI-2 are shown in Table III. 
These are compared with the values obtained with NL-SH and NL3. The SVI
interactions describe the experimental data \cite{Otten.89} on nuclei well. An 
improvement on the charge radii of Pb isotopes over those of NL3 can be seen.

\begin{table}
\caption{\label{tab:table3} The $rms$ charge radius (in fm) obtained with
SVI-1 and SVI-2. The values for NL-SH and NL3 are also shown.
The $rms$ deviation ($\delta$) of theoretical values from the experimental 
data is shown at the bottom of the table.}
\begin{ruledtabular}
\begin{tabular}{c c c c c c c c}
&   Nucleus   & SVI-1 & SVI-2 & NL-SH &  NL3  && ~exp. \\    
\hline 
& $^{16}$O    & 2.698 & 2.700 & 2.699 & 2.728 && ~2.730 \\
& $^{40}$Ca   & 3.438 & 3.442 & 3.452 & 3.470 && ~3.450 \\
& $^{48}$Ca   & 3.451 & 3.457 & 3.452 & 3.470 && ~3.450 \\
& $^{90}$Zr   & 4.277 & 4.285 & 4.289 & 4.287 && ~4.258 \\
& $^{116}$Sn  & 4.593 & 4.601 & 4.599 & 4.599 && ~4.626 \\
& $^{124}$Sn  & 4.644 & 4.652 & 4.651 & 4.661 && ~4.673 \\
& $^{208}$Pb  & 5.501 & 5.508 & 5.509 & 5.523 && ~5.503 \\
& $^{214}$Pb  & 5.562 & 5.568 & 5.562 & 5.581 && ~5.558 \\
\hline 
& $\delta$    & 0.021 & 0.019 & 0.021 & 0.020 &&  \\
\end{tabular}
\end{ruledtabular}
\end{table}

We have calculated the ground-state properties of the isotopic chains 
of Sn and Pb. Especially, the chain of Sn isotopes offers experimental 
binding energies over the whole range from the doubly magic nucleus 
$^{100}$Sn to the doubly magic $^{132}$Sn, thus encompassing the 
space between the two magic numbers $N=50$ and $N=82$. The 
difference $\Delta E$ of binding energy of nuclei obtained with RMF+BCS 
calculations from the experimental value is shown for Sn and 
Pb isotopes in Fig. 1. For the BCS pairing, neutron pairing gaps have 
been obtained from the experimental masses of neighbouring nuclei.

The problem of the arches and a predominance of shell energy at the magic
numbers is well-known. It pervades both the microscopic theories 
as well as macroscopic-microscopic mass formulae. Viewing the results 
for Sn isotopes in  Fig. 1(a), one can notice unambiguous arches at 
the two magic numbers especially with the force NL3. With exception 
of the region near $^{100}$Sn, both SVI-1 and SVI-2 describe the data well 
for nuclei including those near $^{132}$Sn ($N=82$). For nuclei in 
the vicinity of $^{100}$Sn ($N=50$) SVI-1 and SVI-2 show a significant 
improvement over the results of NL3. The $rms$ deviation from the experimental
data is 1.04 MeV and 1.07 MeV for SVI-1 and SVI-2, respectively. In
comparison, it is 1.83 MeV for NL3. This indicates a significant improvement
in the binding energies for Sn isotopes with SVI-1 and SVI-2, especially
near the magic numbers $N=50$ and $N=82$. The arch-like behaviour 
with SVI-1 and SVI-2 is reduced considerably.

This pattern is also visible for the isotopic chain of Pb in Fig. 1(b). 
Both SVI-1 and SVI-2 exhibit a significant improvement in the binding
energies over NL3. The $rms$ deviation of the theoretical values with
SVI-1 and SVI-2 is 0.88 MeV and 0.69 MeV, respectively. This is much
smaller than the corresponding value of 1.82 MeV with NL3 for the Pb
isotopes. Thus, SVI interactions provide a better description 
of the binding energies of the Pb isotopes. In comparison, NL3 values 
overestimate the data near the magic number and gives a well-formed arch 
about the magic number. With NL3 divergences of the binding energies 
near the magic number are displayed strongly as has been observed also for the 
Sn isotopes above. It is a matter of further investigation as to what ingredients 
in the RMF theory would lead to divergences or a lack thereof at shell closures.

%
\begin{figure}
\vspace{0.5cm}
\resizebox{0.80\textwidth}{!}{%
  \rotatebox{0}{\includegraphics{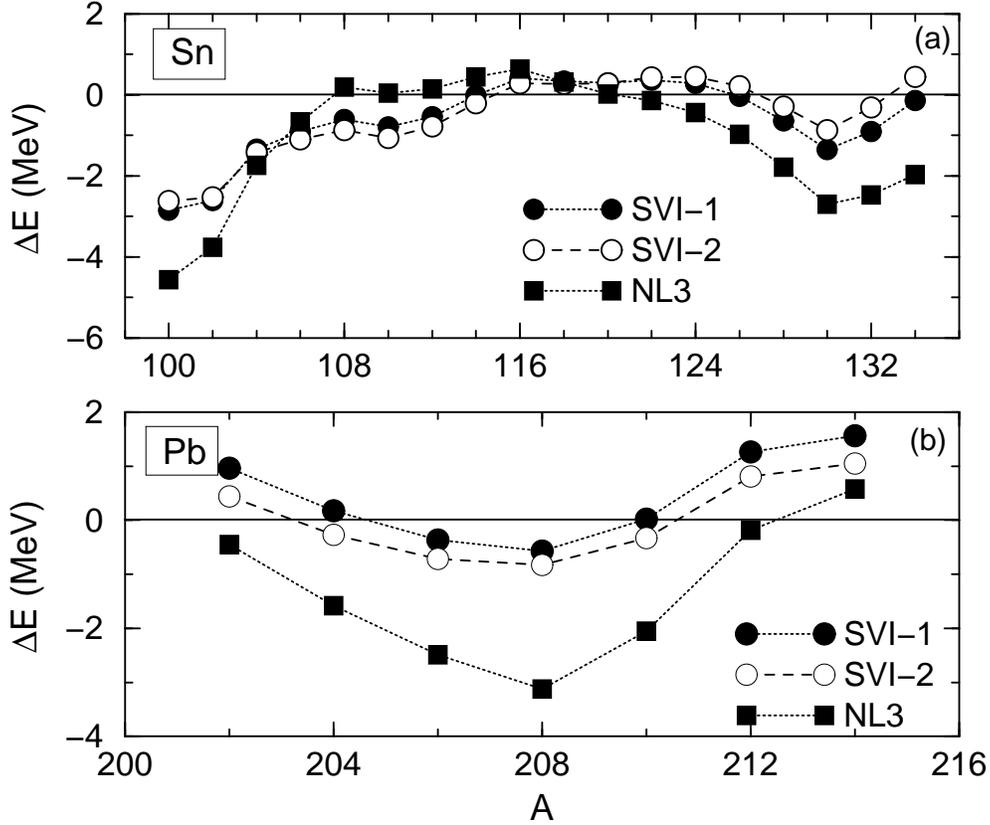}}
}
\caption{Binding energies for (a) Sn and (b) Pb isotopes with SVI-1 and 
SVI-2 and compared with the experimental data. The results obtained 
with NL3 are also shown.}
\label{fig:1}       
\vspace{-0.5cm}
\end{figure}

%
\begin{figure}
\vspace{0.5cm}
\resizebox{0.60\textwidth}{!}{%
  \rotatebox{0}{\includegraphics{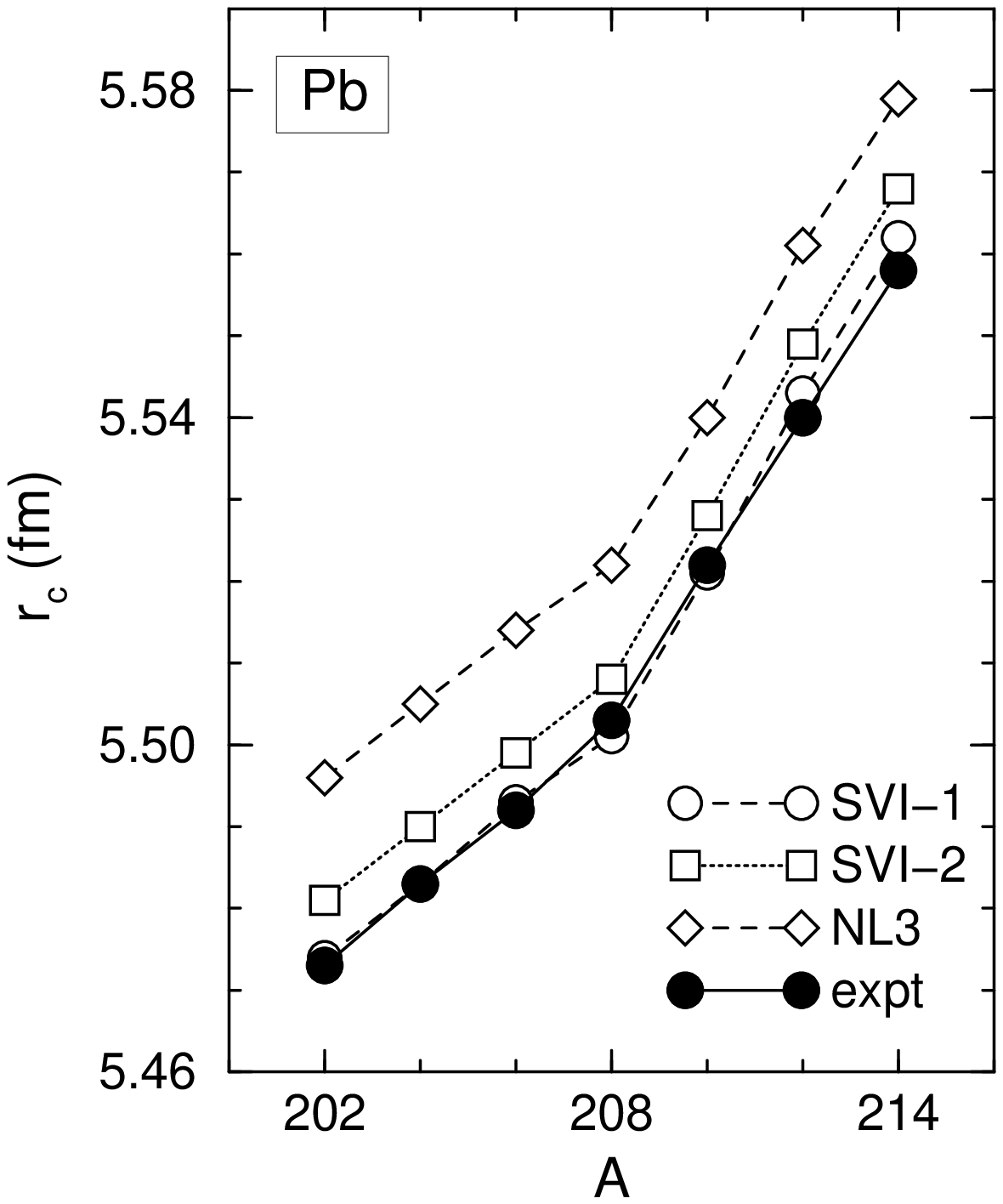}}
}
\caption{Charge radii of Pb isotopes obtained with SVI-1, SVI-2 and NL3 and
compared with the experimental data.}
\label{fig:2}       
\vspace{-0.5cm}
\end{figure}

The charge radii of Pb isotopes and the anomalous kink in charge radii
represent a characteristic feature related to shell structure
of nuclei. It was shown that the RMF theory with NL-SH reproduced
the anomalous kink successfully \cite{SLR.93}. This feature has since
been demonstrated by all the Lagrangian models in the RMF theory. 
In order to discern the behaviour of SVI in this respect, we show 
in Fig. 2 the charge radii of Pb isotopes calculated with SVI-1 and SVI-2.
The force SVI-1 reproduces the experimental data \cite{Otten.89} over the 
range of Pb isotopes very well. In comparison, SVI-2 values overestimate 
the data only slightly. The set NL3, on the other hand, overestimates
the charge radii of all the Pb isotopes significantly.

In order to test the applicability of the new model for nuclei away
from the line of $\beta$-stability, we have performed axially deformed
RMF calculations for a number of nuclei. The nuclei encompass representative
cases from $^{36}$Si to $^{196}$Pt, which includes nuclei from medium
masses through the rare-earth region to higher masses. The results of
calculations with SVI-1 and SVI-2 are shown in Table IV. 
The experimental data on the binding energies are
taken from the recent high-precision mass measurements on 
Si \cite{Jurado.07}, Sr \cite{Sikler.05} and Mo \cite{Hager.06}. The binding
energy of other nuclei has been taken from the 2003 compilation of atomic 
masses \cite{Audi.03}. 

Both SVI-1 and SVI-2 provide an excellent description of the experimental 
binding energies of nuclei over a large range of atomic mass. The difference 
between the predictions of the two sets are small. For a few cases SVI-1 
provides a better description whereas for a few others the SVI-2 does better. 
The $rms$ deviation of the binding energies for both the sets amounts to 
$\sim$ 0.62 MeV. 

\begin{table}
\caption{The binding energy (in MeV) and quadrupole deformation $\beta_2$ 
(in parentheses) of nuclei away from the stability line calculated 
with SVI-1 and SVI-2. The experimental values (exp.) where available
are shown for comparison.}

\begin{ruledtabular}
\begin{tabular}{l l l l l l }
& Nucleus     & ~~~SVI-1 & ~~~~~SVI-2 & ~~exp.      \\    
\hline 
& $^{36}$Si      & $-$292.8 ($-$0.03) & $-$292.7 (0.03)  & $-$292.0  \\
& $^{38}$Si      & $-$300.6 (0.28)    & $-$300.2 (0.28)  & $-$299.9  \\
& $^{40}$Si      & $-$306.9 (0.35)    & $-$306.4 (0.35)  & $-$306.5  \\
& $^{40}$S       & $-$333.2 (0.25)    & $-$332.8 (0.25)  & $-$333.2  \\
& $^{76}$Ni      & $-$632.9 (0.0)     & $-$632.3 (0.0)   & $-$633.1  \\
& $^{86}$Sr      & $-$748.5 (0.0)     & $-$748.5 (0.0)   & $-$748.9  \\
& $^{88}$Sr      & $-$768.1 (0.0)     & $-$768.0 (0.0)   & $-$768.5  \\
& $^{110}$Mo     & $-$918.6 ($-$0.23) & $-$918.5 ($-$0.23) & $-$919.5  \\
& $^{120}$Xe     & $-$1008.1 (0.28)   & $-$1008.4 (0.27) & $-$1007.7 (0.22)\\
& $^{140}$Xe     & $-$1161.3 (0.10)   & $-$1161.2 (0.10) & $-$1160.7 (0.11)\\
& $^{154}$Sm     & $-$1267.4 (0.31)   & $-$1267.5 (0.30) & $-$1266.9 (0.34)\\
& $^{164}$Dy     & $-$1337.9 (0.35)   & $-$1338.1 (0.35) & $-$1338.0 (0.35)\\
& $^{168}$Er     & $-$1365.3 (0.34)   & $-$1365.5 (0.34) & $-$1365.8 (0.34)\\
& $^{174}$Yb     & $-$1406.7 (0.31)   & $-$1407.0 (0.31) & $-$1406.6 (0.33)\\
& $^{190}$W      & $-$1510.5 (0.19)   & $-$1510.7 (0.20) & $-$1509.9  \\
& $^{196}$Pt     & $-$1552.2 (0.12)   & $-$1552.5 (0.12) & $-$1553.6 (0.13)\\
\end{tabular}
\end{ruledtabular}
\end{table}

The original Walecka model (linear) \cite{SW.86} with the nucleon-meson 
couplings of $\sigma$ and $\omega$ mesons has been instructive for 
achieving saturation of nuclear matter. With the inclusion of nonlinear 
scalar self-couplings, the saturation is achieved at nuclear matter 
properties viz., the compressibility within the acceptable range 
(cf. nuclear matter properties in Table I with e.g. NL-SH and NL3).
Here the incompressibility of nuclear matter $K$ denotes the cardinal
point on the saturation curve. The incompressibility $K$ plays an important 
role in determining the breathing-mode giant monopole resonance (GMR) energies.
It is thus important that an acceptable Lagrangian model should be able to 
describe the GMR data.
\noindent
\begin{table}[h*]
\caption{The breathing mode GMR energies obtained with constrained 
GCM calculations using SVI-1 and SVI-2. The experimental 
data  \cite{Young.04,Sharma.88} are also shown.}
\begin{ruledtabular}
\begin{tabular}{c c c c c c}
  Nucleus    & SVI-1  & SVI-2  & NL3  &&  exp. \\    
\hline 
$^{90}$Zr    & 17.2   & 17.5   & 16.9 && $17.81 \pm 0.30$ \\
$^{120}$Sn   & 15.2   & 15.4   & 15.0 && $15.52 \pm 0.15$ \\
$^{208}$Pb   & 13.3   & 13.5   & 13.0 && $13.96 \pm 0.28$ \\
\end{tabular}
\end{ruledtabular}
\end{table}

For a comparative analysis of the Lagrangian models involved, we have 
carried out constrained Generator Coordinate Method (GCM) calculations
\cite{Sto.94} for the isoscalar GMR mode for a set of nuclei. 
The nuclei included are $^{90}$Zr, $^{120}$Sn and $^{208}$Pb. 
The results of GCM calculations are shown in Table V.  

The set SVI-1 with $K=264$ MeV underestimates the experimental values 
for $^{90}$Zr and $^{208}$Pb by more than $\sim$ 0.5 MeV, whereas for
$^{120}$Sn, the disagreement with the datum is nominal. In comparison,
SVI-2 with $K=272$ MeV gives GMR energies for $^{90}$Zr and $^{120}$Sn, 
which are very close to the experimental data. For $^{208}$Pb, its values
is slightly smaller than the experimental one. Comparatively, NL3 with
$K=272$ MeV gives values which are systematically smaller than those of
SVI-2. This difference in predictions of a nonlinear scalar Lagrangian
from those of a scalar-vector Lagrangian points to some subtle differences
in finite effects, possibly surface, of the two models. A comparison 
between the two sets of SVI shows that SVI-2 is commensurate with
the ground-state properties of finite nuclei as well as with the
GMR energies. 

In conclusion, we have constructed a Lagrangian model without nonlinear
scalar self-couplings in the RMF theory. Incorporating scalar-vector
interaction terms in the linear RMF Lagrangian, the Lagrangian model SVI 
has been developed. We have obtained parameter sets SVI-1 and SVI-2 in 
the framework of the new model. It is shown that both SVI-1 and SVI-2 
provide a good description of the ground-state properties of nuclei 
along the stability line as well as for nuclei far away from it. 
With an incompressibility of nuclear matter $K \sim 272$ MeV, 
SVI-2 is able to reproduce the breathing-mode GMR energies on key 
nuclei well. Thus, the model SVI without nonlinear interactions becomes 
viable for finite nuclei and nuclear matter. 

I thank Prof. Lev Savushkin for useful discussions. This work is supported 
by the Research Administration Project No. SP07/03 of Kuwait University.

\end{document}